# Phonon-derived ultrafast relaxation of spin-valley polarized states in MoS$_2$

*Dongbin Shin, Hosub Jin, Noejung Park*[*]

Department of Physics, Ulsan National Institute of Science and Technology (UNIST), 50 UNIST, Ulsan 44919, South Korea

*E-mail: noejung@unist.ac.kr *Tel: +82-52-217-2939; Fax: +82-52-217-2019*

**The valley degree of freedom and the possibility of spin-valley coupling of solid materials have attracted growing interest, and the relaxation dynamics of spin- and valley-polarized states has become an important focus of recent studies. In spin-orbit-coupled inversion-asymmetric two-dimensional materials, such as $MoS_2$, it has been found that the spin randomization is characteristically faster than the time scales for inter- and intra-valley scatterings. In this study, we examined the ultrafast non-collinear spin dynamics of an electron valley in monolayer $MoS_2$ by using real-time propagation time-dependent density functional theory. We found that the spin precession of an electron in the valley is sharply coupled with the lowest-lying optical phonon that release the in-plane mirror symmetry. This indicates that the spin randomization of $MoS_2$ is mainly caused by spin-phonon interaction. We further suggest that flipping of spins in a spin-orbit-coupled system can be achieved by the control over phonons.**

The controlled switching between distinct quantum mechanical states of solid materials is a key component of electronic and information technologies[1]. The charge and spin degrees of freedom have been the main focus of these technologies over the last half century, but recently a new type of degree of freedom, *i.e.* distinct momentum states among degenerate energy multiplet or the so-called valley degree of freedom, has emerged[2-5]. Optoelectronic method has been introduced to achieve the polarization into a specific spin or valley state, and the device structure to measure the valley Hall effect, analogous to the charge or spin Hall effect, has been explored[4-7]. Although the valley degree of freedom can be considered for quite broad classes of materials[8,9], two-dimensional (2D) hexagonal lattices have recently received the most focused attention[1]: graphene and transition-metal dichalcogenides (TMDC) are known to have conical Dirac valleys at the vertices (K and K`) of their hexagonal Brillouin zones. When spin orbit coupling (SOC) arises in the inversion-asymmetric trigonal structure of a TMDC, the valley states exhibit a particular spin state depending on their momentum vector, leading to an intrinsic spin-valley coupled state[2,3,10].

The most prototypical 2D material in the perspective of spin-valley polarization is monolayer $MoS_2$. In contrast to the indirect band gap structures of stacked multiple layers, the isolated monolayer $MoS_2$ has direct band gaps at K and K` points in the Brillouin zone[11]. The valence band maxima (VBMs) of those valleys exhibit sizable spin splitting, whereas the conduction band minima (CBMs) are almost degenerate. Recently, Mak *et al.* and Zeng *et al.* demonstrated that, owing to this spin-valley coupled VBM structure, a circularly polarized photon can excite spin-polarized electrons only at a particular valley of monolayer $MoS_2$ [2,3]. Once the potential of $MoS_2$ in valleytronics operations was identified, microscopic understanding of the relaxation dynamics of either the spin- or valley-polarized is now considered an important issue[10,12,13]. It was reported that, compared with the time scales for the relaxation of valley-polarized state, the spin randomization time is substantially short. Such spin dynamics in MoS$_2$ is thought to be even faster than the spin dynamics in similar other materials. The previous pump-probe measurement using circularly polarized light suggested that the time for electron spin randomization, inter-valley scattering, intra-valley scattering, and electron-hole recombination are around 60 fs, 1 ps, 25 ps, and 300 ps at 78 K, respectively[10].

In this study, we investigated the ultrafast spin relaxation in an electron valley of monolayer MoS$_2$ by performing real-time propagation time-dependent density functional theory (rtp-TDDFT) calculations. Although phonons have been conjectured to have a vital role, no clear understanding has been achieved of the effect of phonon on the spin-valley polarized monolayer MoS$_2$ [13]. Here, through the full ***ab initio*** simulation of lattice and non-collinear spinor dynamics, we determined that the ultrafast relaxation of spins in the polarized valley state arises from the coupling between the spin state of CBM and the mirror-asymmetric optical phonon: the $E''$ optical phonon derives the spin precession through the effective magnetic field carried by the SOC term.

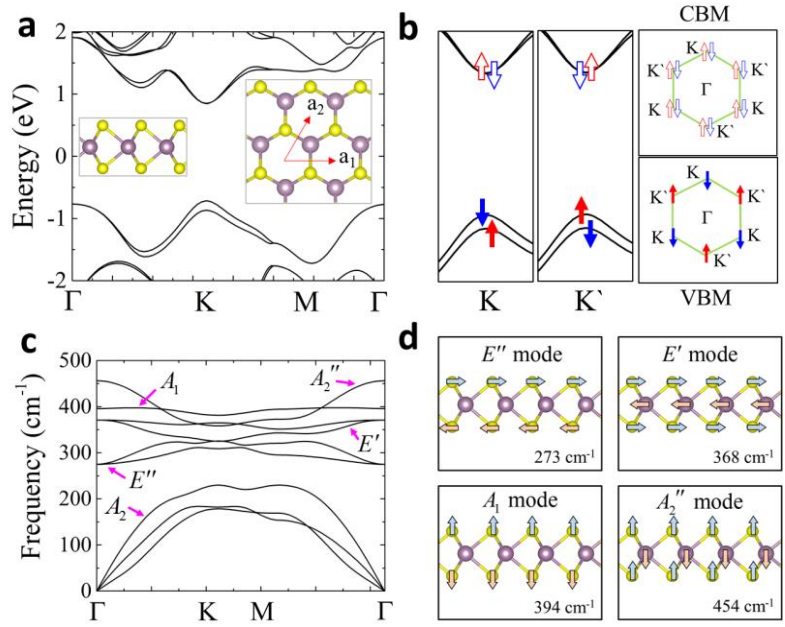

**Figure 1 | Electronic structure and phonon dispersion of monolayer MoS$_2$. a**, Band structure of monolayer MoS$_2$. **b**, Spin-polarized bands near K and K` (left), and spin structure of the VBMs and CBMs in the Brillouin zone (right). **c**, Phonon dispersion of monolayer MoS$_2$. **d**, The four optical phonon modes at q = (0, 0, 0). The insets in (**a**) are the side view (left) and top view (right) atomic geometries of monolayer MoS$_2$. The purple and yellow balls in (**a**) and (**d**) are Mo and S atoms, respectively. The red arrows in the inset on the right in (**a**) are the lattice vectors of the triangular cell. The red and blue arrows in (**b**) describe the spin-up or spin-down states, where solid and outlined arrows represent occupied and unoccupied states, respectively.

The intriguing spin-valley properties of monolayer MoS$_2$ should be discussed in relation with its geometry. As shown in the inset in **Figure 1a**, monolayer MoS$_2$ constitutes the honeycomb structure with the mirror plane at the central Mo layer. The SOC interaction of the Mo atomic orbitals in this inversion-asymmetric lattice caused the spin splitting over the Brillouin zone except the mirror symmetry-protected degenerate points. The splitting of VBMs amount to 156 meV near K and K`, as shown in **Figure 1a and 1b,** while the CBMs experience a relatively small splitting of 3 meV[14]. The presence of central mirror plane in the point group of $D_{3h}$ constrains the spin of each band to align in the out-of-plane direction[2,5,15].

Before we describe our rtp-TDDFT simulation results, we need to introduce the phonons of monolayer $MoS_2$, as plotted in **Figure 1c and 1d**. Above the acoustic phonons ($A_2$), there are four optical phonon modes: mirror symmetry breaking in-plane motion of S ($E''$), mirror symmetry preserving in-plane alternating motion of S and Mo ($E'$), mirror symmetry preserving out-of-plane movement of S ($A_1$), and mirror symmetry breaking out-of-plane movement of S and Mo ($A_2''$) have phonon frequencies of 273 cm$^{-1}$, 368 cm$^{-1}$, 394 cm$^{-1}$, and 454 cm$^{-1}$, respectively, at q = (0, 0, 0), as shown in **Figure 1d**. Note that the regular hexagonal structure is also broken by $E'$ and $E''$ phonon modes.

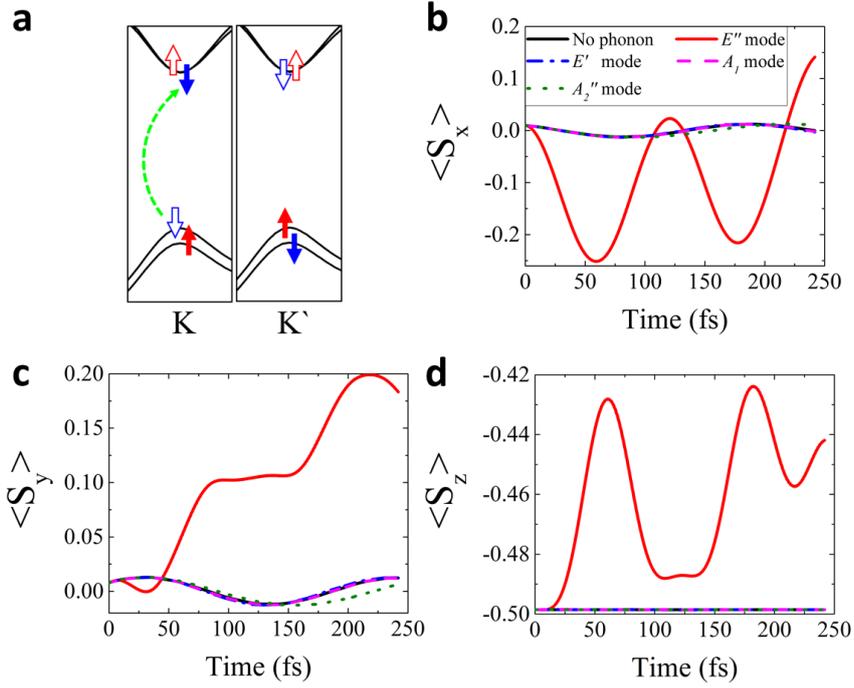

**Figure 2 | Rtp-TDDFT result of spin dynamics of electron on conduction band minimum at the K point. a**, Schematic diagram of the electron excitation at K point. **b-d**, Time evolution of the Cartesian spin components of the excited electron when each one of four optical phonon was selectively turned on and without initial phonon mode. The green curved dashed arrow in (**a**) indicates schematic excitation VBM to CBM spin down state.

We performed non-collinear rtp-TDDFT simulations to explore the effect of coherent phonon on the excited electron at a valley. In rtp-TDDFT calculations, as explained in detail in the computational method section of Supplementary Information, two component Kohn-Sham spinors ($|\psi_{n,\vec{k}}\rangle$) evolve along the time through the time-dependent Kohn-Sham equation:

$$i\hbar \frac{\partial}{\partial t}|\psi_{n,\vec{k}}\rangle = \left( -\frac{\hbar^2}{2m}\nabla^2 + \sum_{\lambda} \hat{v}_{pp}(\vec{r}-\vec{R}_{\lambda}(t)) + v_{Hxc}[\rho(t)] + \hat{v}_{SOC} \right)|\psi_{n,\vec{k}}\rangle \quad \text{(eq. 1)},$$

where atomic geometry ($\vec{R}_\lambda(t)$) is also updated following Ehrenfest dynamics. The spin and the charge density at time $t$ were derived from the time-evolved Kohn-Sham orbital: $\vec{S}_{n,\vec{k}}(t) = \langle \psi_{n,\vec{k}}(t) | \frac{\hbar}{2} \vec{\sigma} | \psi_{n,\vec{k}}(t) \rangle$ and $\rho(t) = \sum_{n,\vec{k}} \psi^*_{n,\vec{k}}(t) \psi_{n,\vec{k}}(t)$. **Figure 2** shows time profiles of the excited electron's spin state at the conduction band edge of the K valley. As an initial configuration of electronic structure (at $t=0$) the spin down electron was relocated from the VBM to the CBM, as depicted in **Figure 2a**. This corresponds to an excitation induced by absorption of right-handed circularly polarized light. The circularly dichroism of monolayer MoS$_2$ is depicted in **Supplementary Figure 1** evaluated by rtp-TDDFT calculation.

Time evolutions of the Cartesian spin components, as results of rtp-TDDFT calculation, are presented in **Figures 2b**, **2c**, and **2d**. To simulate the effect of phonon on the spin dynamics, atoms departed from the equilibrium positions with the initial velocity in eigenvector direction of a particular phonon with the total kinetic energy of 12.2 meV per the unit cell. It is greatly noteworthy that only the $E''$ phonon mode appreciably derives spin motions. The effect of the other phonons, depicted by dash dot blue ($E'$), dash pink ($A_1$), and dot green lines ($A_2''$) in **Figures 2b**, **2c**, and **2d**, are negligible. As a reference we also calculated the spin dynamics on the fixed equilibrium geometry and confirmed that the spin dynamics deduced from the time-dependent spinor have negligible time variation. The details of initial condition of coherent phonon mode in rtp-TDDFT calculation is discussed in **Supplementary Information**.

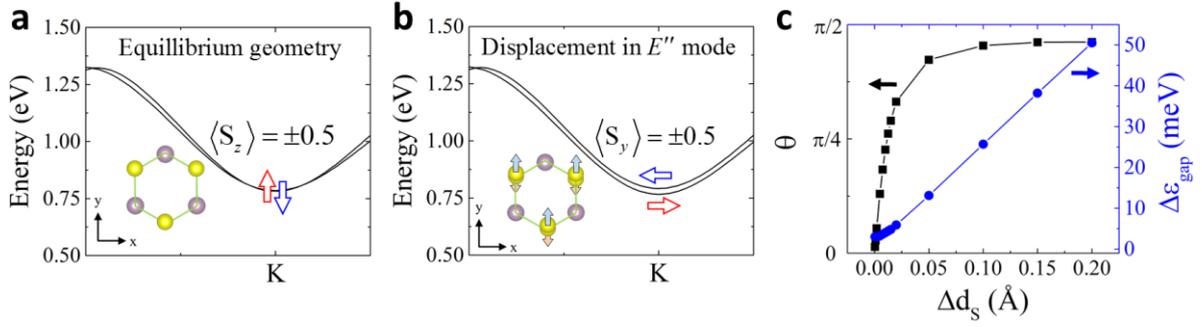

**Figure 3 | Spin-resolved electronic structure near the CBM of K point as a result of static lattice deformation along $E''$ phonon mode in $\hat{y}$ direction. a-b**, Conduction band minimum near the K point of (**a**) the equilibrium geometry and (**b**) the atomic structure displaced by 0.1 Å in the eigenvector direction of the $E''$ phonon mode. **c**, Variations in the inclination spin angle defined as $\theta = \arccos\left(\langle S_z \rangle / \left|\langle \vec{S} \rangle\right|\right)$ and the spin splitting ($\Delta\varepsilon_{gap}$) with respect to the displacement of the atomic geometry in the eigenvector direction of the $E''$ phonon mode.

The critical role of the $E''$ phonon mode, as explained in the previous paragraph, can be attributed to the strong SOC. To understand its effect more explicitly, we investigated the variation in the electronic structure with respect to static displacements along eigenvectors of the phonon mode. **Figure 3a** and **3b** show the spin-resolved band structure near CBM at K valley of the equilibrium geometry and the displaced geometry in the eigenvector direction of the $E''$ mode, respectively. As explained above, in the equilibrium geometry, the spin at the CBM is polarized in the perpendicular direction ($\hat{z}$) as depicted in **Figure 3a**. With the displacement along the $E''$ mode vector in $\hat{y}$ direction, the spins are polarized in $\hat{y}$, as depicted in **Figure 3b.** The $E''$ mode is doubly degenerate at $\Gamma$, thus the linear combination of the two eigenvector can be chosen to any direction within the plane. Details are explained in **Supplementary Table 1**. Similar band structures of the geometries statically displaced along the eigenvector of the other phonons, such as $E'$, $A_1$, and $A_2''$, are summarized in **Supplementary Figure 2**. It is noteworthy that the in-plane mirror symmetry, as in the equilibrium geometry, is preserved over the $E'$, $A_1$, and $A_2''$ phonon modes, and thus the spins are constrained to the perpendicular direction. For all these cases, the splitting between the spin-up and spin-down state is as small as 3 meV, as depicted in **Figure 3a** and also in **Supplementary Figure 2**. However, the displacement in the eigenvector direction of the $E''$ mode gave rise to quite distinctive features, as shown in **Figure 3b**: The in-plane mirror symmetry is lifted and the spin lies in the plane direction with a larger spin splitting. Various amounts of distortions were tested between the equilibrium structure and the maximally shifted one in which S atom is displaced by 0.2 Å from the equilibrium. The spin splitting gap ($\Delta\varepsilon_{gap}$) of the CBM at the K point and the inclination angle of the spin vector, defined by ($\theta = \arccos\left(\langle S_z \rangle / |\langle \vec{S} \rangle|\right)$), are plotted in **Figure 3c**. This result demonstrates that the displacement as large as 0.05Å along the $E''$ mode in $\hat{y}$ direction readily induces the rotation of the spin polarization from the perpendicular direction to the in-plane one.

Aforementioned effect of lattice distortion on the spin direction can be described by the effective magnetic field generated through the SOC. The splitting between the spin-up and down state in the perpendicular direction of the equilibrium geometry, as presented in **Figure 3a**, can be associated with the Zeeman splitting induced by the magnetic field, namely $\vec{B} = B_0 \hat{z}$. Lattice displacements along with the emergence of phonon modes can induce additional magnetic field, as explained above. For example, as the lattice displaces along the $E''$ mode in $\hat{y}$ direction, the spin of the electron at the CBM polarizes in $\hat{y}$ direction, as shown in **Figure 3b**, which can be associated with the magnetic field of $\vec{B} = B_{ph}\hat{y}$. It is noteworthy that the same $E''$ phonon mode, through the combination of the doubly degenerate eigenvector, can be directly differently within the plane. The features of the spin polarizations along with the $E''$ phonon mode turned on various in-plane directions are discussed in **Supplementary Information**: for example, when the lattice was displaced to $\hat{x}$ direction, the spin-polarization vectors become to align in $\hat{x}$ direction.

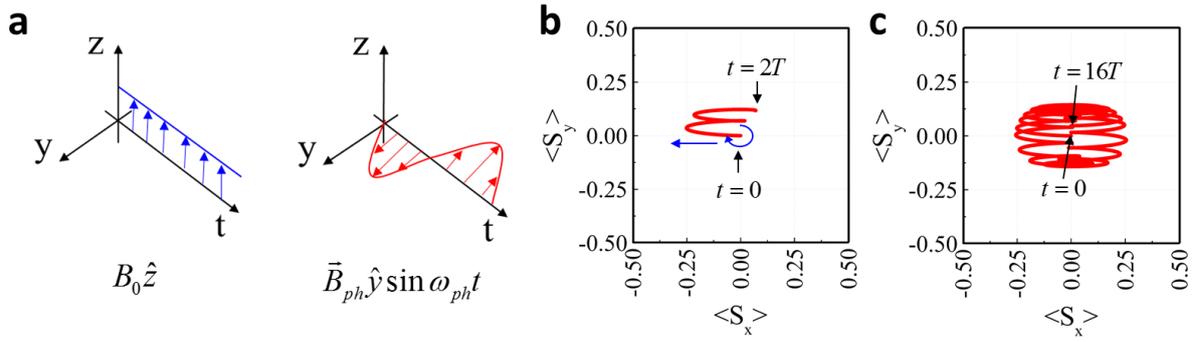

**Figure 4 | Time evolution of spinor calculated by the simplified Hamiltonian with two component of magnetic field. a**, Schematics of the two components of the magnetic field which is (**a**) static in $\hat{z}$ direction (left) and the oscillating $\hat{y}$ direction (right). **b-c**, Spin-trajectories up to (**b**) $t = 2T$ and (**c**) $t = 16T$. The curved blue arrow and the straight blue arrow in (**b**) depict the torque on the spin near $t = 0$.

The actual effect of dynamic lattice introduces a periodically oscillating magnetic field. For example the $E''$ phonon mode in $\hat{y}$ direction, as described in **Figure 3**, can be associated with the magnetic field $\vec{B} = B_0 \hat{z} + B_{ph}(t)\hat{y}$, where the latter periodically oscillates with the phonon frequency: $B_{ph}(t) = B_{ph}(t + 2\pi/\omega_{ph})$. Prior to introducing detailed results of ***ab initio*** calculations, here we first examine the spin dynamics by considering simplified model Hamiltonian featured by this two-component magnetic fields: one is static in the perpendicular direction ($\hat{z}$) and the second is oscillating in the plane. With $\vec{B}(t) = B_0\hat{z} + B_{ph}\hat{y}\sin(\omega_{ph}t)$, as depicted in **Figure 4a**, the model Hamiltonian can be written as follows.

$$\hat{H}(t) = \frac{e}{m}\hat{\vec{S}}\cdot\vec{B} = \varepsilon_0\hat{\sigma}_z + \varepsilon_{ph}\hat{\sigma}_y\sin(\omega_{ph}t), \text{ where } \varepsilon_0 = \frac{e\hbar}{2m}B_0, \ \varepsilon_{ph} = \frac{e\hbar}{2m}B_{ph} \quad \text{(eq. 2)}.$$

Based on the obtained spin splitting of the CBM state at K, that is 3 meV, as presented in **Figure 3a**, we take the value $\varepsilon_0 = 1.5$ meV, and $\omega_{ph}$ is taken as that of $E''$ phonon mode, namely $\omega_{ph} = \frac{2\pi}{122} fs^{-1}$. The variable $\varepsilon_{ph}$ is defined by the magnitude of the magnetic field which is characterized by the amplitude of the lattice oscillation or the phonon occupation number. We calculated the time propagation of the spinor assuming the initially spin-down state.

$$|\psi(t+\Delta t)\rangle = \exp\left(-\frac{i}{\hbar}\hat{H}(t)\Delta t\right)|\psi(t)\rangle,$$
$$\text{with } |\psi(t=0)\rangle = \begin{pmatrix} 0 \\ 1 \end{pmatrix} \quad \text{(eq. 3)}.$$

For $\varepsilon_{ph} = 3\varepsilon_0$ the trajectory of tip of spin vector are plotted up to $t = 2T = 4\pi/\omega_{ph}$, as presented in **Figure 4b**. It is greatly noteworthy that, opposed to the simple Larmor rotation derived by the static magnetic field, the spin vector is not recovered to its original position. This complicated behavior can be explained by the structure of the torque on the spin:

$d\vec{S}/dt = \vec{S} \times \vec{B}$. At around $t \approx 0$, the $S_x$ develops in proportional to $-\frac{\hbar}{2}B_{ph}(t)$, and the emerging $x$ component contributes to the development of the $y$ component in proportional to $-S_x(t)B_0$. Thus the tip shifts along the $y$ direction, as depicted in **Figure 4b**. In this dynamics, the magnitude of $B_0$ and $B_{ph}$, and the frequency of phonon ($\omega_{ph}$) are intimately interrelated. The same trajectory over a longer time, up to $t = 16T$, is shown in **Figure 4c**. There are more results of spin dynamics simulations evaluated by model Hamiltonian time-propagation with various $B_{ph}$ values in **Supplementary Figure 4**.

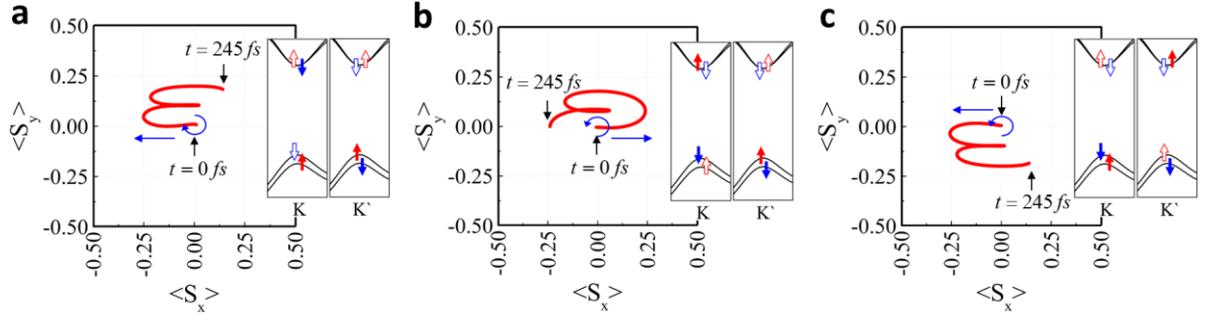

**Figure 5 | Spin dynamics calculated by rtp-TDDFT with different initial excited state configurations. a-c**, Spin-trajectories in the presence of the $E''$ phonon mode for (**a**) the spin-down state at K, (**b**) the spin-up state at K, and (**c**) the spin-up state at K`. The straight and curved blue arrows indicate the schematic directions of the torque on the spin at $t = 0$ fs. The insets to the right in (**a**), (**b**), and (**c**) describe the initial excited state configurations of the spin states and their occupations at $t = 0$ fs. The solid and outlined arrows in inset of (**a**), (**b**), and (**c**) indicate the occupied and unoccupied states respectively.

The trajectory of spin of excited electron at the CBM calculated by the rtp-TDDFT

corresponds well to the result of the model Hamiltonian. Here we compare the *ab initio* spin dynamics calculated with different initial excited state conditions: the spin-down electron at the K valley, the spin-up electron at the K valley, and the spin-up electron at the K`. With given phonon induced magnetic field, the spin-down electron processes in the opposite way to the spin-up electron, as we observe in the comparison with **Figure 5a** and **5b**. More intriguing result can be obtained from the comparison between the spin-up electron at K valley and the same spin at K` valley, as presented in **Figure 5c** and **5d**. The rotation generated by the static field ($B_0\hat{z}$) is the same for both electrons. But the effect of phonon on the electron in K is opposite to that on K`. This is obviously due to the structure of the SOC: $\hat{V}_{SOC} = \frac{e\hbar}{4m^2c^2}\hat{\vec{\sigma}} \cdot \left(\vec{\nabla}V \times \hat{\vec{p}}\right)$. The given phonon vector induces the same change in the scalar potential ($\vec{\nabla}V$), but the opposite momentum vector produce the oppositely directed magnetic field. We also discuss spin dynamics of excited electron in valley with different direction of E`` phonon mode, initial kinetic energy, and various conditions in **Supplementary Figure 4, 5, and 6**, respectively.

In conclusion, we have determined the mechanism of spin relaxation of the electron valley state of monolayer $MoS_2$. By performing non-collinear rtp-TDDFT calculations with a SOC combined with the Ehrenfest dynamics of lattice vibration, we found that the lowest-lying optical phonon mode, namely the $E''$ phonon mode, solely derives the spin rotation of the electron valley state, while the other phonons are almost irrelevant. This suggests that the emergence of many incoherent phases of that phonon can randomize the spins within the time scales of hundreds femtoseconds, no matter how well the spins are polarized.

## Methods

**Computational details** To calculate the ground state electronic structure and phonon dispersion we used Quantum ESPRESSO package[16]. For the non-collinear Kohn-Sham wavefunctions with SOC interaction, the plane-wave basis set with 30 Ry energy cut-off, Perdew-Burke-Ernzerhof type generalized gradient approximated exchange-correlation functional, and projector augmented wave (PAW) method with full-relativistic potential were used[17]. The primitive unit cell with the lattice vector of $a = 3.15$ Å and the vacuum slab of 15 Å vacuum were used to simulate the monolayer $MoS_2$. The whole Bouillon zone was sampled with the grids of 18x18x1 points excluding any symmetric operation.

For time evolution of the non-collinear Kohn-Sham wavefunctions we used the package developed by ourselves[18]. We tested the accuracy by varying a few parameters for the time propagation, and the presented results were calculated by Crank-Nicolson propagator with $\Delta t = 2.42$ attoseconds, which preserve the total energy within $5.3 \times 10^{-5}$ eV over 245 femtoseconds. Instead of the full grid of $18 \times 18 \times 1$, that used for the ground state calculation, we used the $6 \times 6 \times 1$ k-points for the time-evolution of Kohn-Sham states which includes the occupied VBM states and excited CBM states. The additional descriptions of rtp-TDDFT and model Hamiltonian calculations are described in Supplementary Information.

## Acknowledgements

This work was supported by the programs through the National Research Foundation (NRF) of Korea.